\pdfoutput=1
\documentclass[11pt,a4paper]{article}
\usepackage{amsmath,amssymb}
\usepackage{graphicx}
\usepackage{cite}
\usepackage{url}

\renewcommand\[{\left[}

\def\sst{\scriptscriptstyle}

\def\beq{\begin{equation}}
\def\eeq{\end{equation}}

\newcommand{\startappendix}{
\setcounter{section}{0}
\renewcommand{\thesection}{\Alph{section}}
\renewcommand{\theequation}{\Alph{section}.\arabic{equation}}}

\newcommand{\Appendix}[1]{
\refstepcounter{section}
\begin{flushleft}
{\Large\bf Appendix: #1}
\end{flushleft}}

\begin{document}

\numberwithin{equation}{section}
\title{{\normalsize  \mbox{}\hfill IPPP/13/65, DCPT/13/130}\\
\vspace{2.5cm}
\Large{\textbf{Inflation and Dark Matter in the Higgs Portal\\
of the Classically Scale Invariant Standard Model}}}

\author{Valentin V. Khoze\footnote{valya.khoze@durham.ac.uk}\\[4ex]
  \small{\em Institute for Particle Physics Phenomenology, Department of Physics} \\
  \small{\em Durham University, Durham DH1 3LE, United Kingdom}\\[0.8ex]
}

\date{}
\maketitle

\begin{abstract}
  \noindent
 We consider a minimal classically scale-invariant extension of the Standard Model.
In this theory, the Higgs mechanism is triggered and the electroweak symmetry breaking is generated radiatively 
by the Coleman-Weinberg sector which is coupled to the SM Higgs. We extend the Higgs portal interactions of the 
theory to include an additional singlet which is also non-minimally coupled to gravity. 
This generates a single-field slow-roll inflation mechanism in the effective 
 field theory formulation which is robust up to Planck scales. Our approach does not require integrating in any
 additional new physics degrees of freedom to unitarise the theory in the sub-Planckian regime where inflation happens.
 As a result,  no large threshold corrections appear in our approach to inflation so that the electroweak
 scale and the SM Higgs mass are not affected. The singlet field responsible for inflation also gives a viable dark matter candidate
 in our model. We also discuss the relation between classical scale-invariance of the effective theory and the possible 
 local scale invariance of the full theory and comment on the naturalness of the electroweak scale. 
\end{abstract}
\thispagestyle{empty}
\setcounter{page}{0}

\newpage

\section{Introduction}\label{sec:intro}

The discovery of the $\simeq 125$ GeV scalar particle \cite{ATLAS:2012gk,CMS:2012gu}
with the properties of the Standard Model Higgs boson together with the so far negative results for searches of 
supersymmetry are pointing to a possibility of a very different model building paradigm based on a minimally
extended Standard Model with manifest classical scale invariance. 
In this paper we will present the first, to the best of our knowledge, realisation of a slow-roll cosmological inflation mechanism 
in a classically scale-invariant BSM theory.

The rational for advocating this `heretical' approach to model building based on classical scale invariance is
as follows.


(1.) First point in favour of (approximate) scale symmetry is a well-known `experimental' fact that
there is just a single occurrence of a non-dynamical scale in the Standard Model (SM) -- the Higgs
mass-squared input parameter $\mu^{2}_{\rm SM}$ -- appearing in the SM Higgs potential,
\begin{equation}
  \label{VH}
  V_{\rm cl}^{\rm SM}(H)\,=\, \mu^{2}_{\rm SM} \,H^{\dagger}H \,+
  \, \frac{\lambda_{\rm H}}{2}\,\left(H^{\dagger}H\right)^{2}\,.
\end{equation}
By selecting a particular negative value for it, 
an expectation value $v \simeq 246~{\rm GeV}$ for the Higgs field, and the
Higgs mass $m_h \simeq 125~{\rm GeV}$ are obtained,
$  \mu^{2}_{\rm SM}\, = -\,\frac{1}{2}\, m_h^2\,=-\,\frac{1}{2}\,\lambda_{\rm H} \, v^2$.
Classical scale invariance, broken by $\mu^{2}_{\rm SM}$, is recovered by re-interpreting this scale
in terms of a vacuum expectation value of a new scalar $\phi$, coupled to the SM via the
Higgs portal interaction, $\lambda_{\rm P}|H|^2|\phi|^{2},$
\begin{equation}
  \label{potentialcoupled1}
  V_{\rm cl}(H,\phi)\,=\, \frac{\lambda_{\rm H}}{2}(H^{\dagger} H)^{2}
   \,+\,\frac{\lambda_{\phi}}{4!}|\phi|^{4}
    \,-\, \lambda_{\rm P}(H^{\dagger}H)|\phi|^{2}\,, \qquad 
  \mu^{2}_{\rm SM}\, = -\,   \lambda_{\rm P} |\langle \phi\rangle |^{2}\,.
\end{equation}  
The classical theory described by \eqref{potentialcoupled1} is scale-invariant, and 
insofar as the appropriate natural value for   $\langle \phi\rangle\ll M_{\sst UV}$ can be generated quantum mechanically,
it will trigger the electroweak symmetry breaking.

\bigskip

(2.) Already 40 years ago in Ref.~\cite{Coleman:1973jx} Coleman and Weinberg discovered
that a massless scalar field $\phi$ coupled to a gauge field
dynamically generates a vacuum expectation value
via dimensional transmutation
of the logarithmically running coupling constants. The $\langle \phi\rangle$ is non-vanishing, calculable in a weakly-coupled theory,
and its value is naturally small (i.e. exponentially suppressed) relative to the UV cutoff of the theory,
\begin{equation}
  \label{run2s}
  \langle \phi\rangle \,  \sim\, M_{\sst UV}\times \exp \left[ -\, \frac{\rm const}{g^{2}_{\sst CW}}
  \right] \, \ll M_{\sst UV}\,,
\end{equation}
where $g_{\sst CW}$ is the gauge coupling of $\phi$. 
In the Appendix we review the Coleman-Weinberg mechanism to recall some of the key expressions for the convenience of the Reader.

\bigskip

In the SM$\times$U(1)$_{\sst CW}$ theory with the classical scalar potential 
\eqref{potentialcoupled1}, the $\phi$-vev in \eqref{run2s} then generates the Higgs vev $v$ and the Higgs mass $m_h$,
as dictated by \eqref{potentialcoupled1},
\begin{equation}
  \label{musmtwo}
   |\langle \phi\rangle|^2 \, =\, \frac{\lambda_{\rm H}}{\lambda_{\rm P}} \, \frac{v^{2}}{2}\, \qquad
 |\langle \phi\rangle|^2 \, =\, 
 \frac{1}{\lambda_{\rm P}} \, 
  \frac{m_h^{2}}{2}   \,,
\end{equation}
and one an set  $v= 246~{\rm GeV}$, $m_{h} = 125~{\rm GeV}$ and $\lambda_{\rm H}=\frac{m_h^{2}}{v^{2}}$. The
U(1)$_{\sst CW}$ theory coupled to the SM via the Higgs portal with the scalar potential \eqref{potentialcoupled1} was first considered in
\cite{Hempfling:1996ht}.

Note that no input mass scales are allowed in the Coleman-Weinberg theory and, in the course of UV renormalisation, the subtraction scheme is 
chosen to set the {\it renormalised masses} at the origin of the field space to zero,
\begin{equation}
  \label{eq:m0}
  m^2|_{\phi=0}\, := \, V^{\prime\prime}(\phi)\bigg|_{\phi=0}=0 \, ,
\end{equation}
where $V$ is the quantum-corrected effective potential. 
Does the masslessness condition \eqref{eq:m0} amount to a fine-tuning?

A theory with no input mass-scales is classically scale invariant~\cite{Meissner:2006zh}. The scale invariance is not exact, 
but neither it is broken 
by an arbitrary amount. The violation of scale invariance is
controlled in quantum theory by the anomaly in the trace of the energy-momentum tensor, or
in other words, by the logarithmic running of dimensionless coupling constants and their dimensional 
transmutation scales which, in the weakly coupled perturbation theory, are much smaller than the UV cutoff.
Bardeen has argued in \cite{Bardeen:1995kv}  that in order to maintain anomalously broken scale invariance, one should 
select a regularisation scheme
which does not explicitly break classical scale invariance (so that it does not introduce positive powers of the UV cutoff scale into renormalised quantities).

In dimensional regularisation, the equation \eqref{eq:m0} is satisfied automatically. Indeed, no power-like  dependences on the cutoff scale appear
  in dimensional regularisation, and in theories like ours, which
  contain no explicit mass scales at the outset, no finite
  corrections to mass terms at the origin of the field space can appear either. In this regularisation which preserves classical
  scale invariance, the theory as it stands is not fine-tuned, at least, not in the technical sense.
  
 If the higher theory above the Planck scale breaks the classical scale invariance of its IR effective theory, through heavy degrees of 
 freedom coupled to the Standard Model,\footnote{ Assuming that the higher theory maintains no symmetry principle which could cancel its heavy contributions to the Higgs mass.}
 then from the perspective of this microscopic theory, the masslessness  condition does amount to a fine-tuning.
 But from the perspective of our sub-Planckian theory used for particle physics computations, there is no fine-tuning,
 and no practical problems arise associated e.g. with cancellations between large scales, etc.
   
Furthermore, once the masslessness condition is enforced at one scale, it holds at all RG scales \cite{Englert:2013gz}.
To summarise, within dimensional regularisation, the renormalisation
condition \eqref{eq:m0} and, more generally, equations,
\begin{equation}
  \label{rencond0}
  \frac{\partial^{2} V(H,\phi)}{\partial H^{\dagger} \partial H}\bigg|_{H=\phi=0}=0\,,\quad
  \frac{\partial^{2} V(H,\phi)}{\partial \phi^{\dagger} \partial \phi}\bigg|_{H=\phi=0}=0\,,
\end{equation}
are self-consistent, contain no fine-tuning in the theory at hand, and are independent of the RG-scale.

\bigskip
The
phenomenology of the SM$\times$U(1)$_{\sst CW}$ model \eqref{potentialcoupled1}  in the context of LHC, future
colliders and low energy measurements was analysed recently in \cite{Englert:2013gz}.
The minimal model has only two
remaining free parameters, the portal coupling, $\lambda_{\rm P}$ and the mass of the second 
scalar eigenstate $m_{h_2}$ (with $m_{h_1}$ being the observed Higgs mass)\footnote{The Higgs field and the CW field 
both have vevs and mix thus the mass of the second 
scalar eigenstate is the relevant physical parameter rather than the mass of the CW scalar. 
This parameter can also be traded for the mass of the CW gauge field $Z'$.}, 
and it was shown 
that the model is perfectly
viable. In particular, the presently available Higgs data 
constrains the portal coupling to be $\lambda_{\rm P} \lesssim 10^{-5}$ on the part of the parameter space where the second scalar is in the region between $10^{-4}$ GeV and 
$m_{h_1}/2$.  For heavier $m_{h_2}$, the portal coupling is much less constrained experimentally, and has a theoretical upper limit of 
$\lambda_{\rm P} \lesssim 10^{-2}$, see Fig.~2 of Ref.~\cite{Englert:2013gz} for more detail.

 Other related studies of classically scale-invariant models can be found in~\cite{Chang:2007ki,Foot:2007as,Foot:2007iy,
 Iso:2009ss,AlexanderNunneley:2010nw,Iso:2012jn,Chun:2013soa,Heikinheimo:2013fta,Hambye:2013dgv,Khoze:2013oga,Carone:2013wla}.

\bigskip

(3.) As the resulting theory has to be valid up to a very large UV scale,
the classical scale invariance provides a powerful principle for the BSM model building. 
No vastly different scales\footnote{Apart from the Planck scale $M$ which is treated separately.} can co-exist in the theory: first, it is difficult to generate a large hierarchy of scales from the same $\langle \phi \rangle$. 
Secondly, the large scales, if they do appear, would ultimately couple to the Higgs and would destabilise it mass.
The BSM theory is a minimal extension of the SM which should address all the sub-Planckian shortcomings of the SM, such as the 
generation of matter-anti-matter
asymmetry, inflation, dark matter, stabilisation of the SM Higgs potential, without introducing scales higher than 
$\langle \phi \rangle$ which itself is not much higher the electroweak scale.

\bigskip

Can one generate the matter-anti-matter asymmetry
of the Universe in these settings?

\medskip

It was shown in Ref.~\cite{Khoze:2013oga} that the classically-invariant 
SM$\times$U(1)$_{\sst CW}$ theory \eqref{potentialcoupled1} where the
Coleman-Weinberg U(1)$_{\sst CW}$  sector is interpreted as the gauged $B-L$ symmetry of the SM, can indeed generate the observed 
value of matter-anti-matter asymmetry. The  $B-L$ model \cite{Mohapatra:1980qe} is an appealing extension of the SM as it  
automatically contains 
three generations of right-handed Majorana neutrinos necessary (needed to cancel the gauge anomaly of  U(1)$_{\sst B-L}$). 
The standard see-saw mechanism generates masses of visible neutrinos and facilitate neutrino oscillations. 
The CW field $\phi$ carries the $B-L$ charge 2 and its vev generates the
Majorana neutrino masses and the mass of the U(1)$_{\sst B-L}$ $Z'$ boson. In the
classically conformal settings, these models were first considered in \cite{Iso:2009ss,Iso:2012jn}. As 
well-known, however, the
standard thermal leptogenesis formalism~\cite{Fukugita:1986hr} requires very heavy Majorana neutrino masses of the order of $10^9$ GeV, 
to generate the observed baryon asymmetry of the Universe. Such heavy neutrino masses go against the grain of 
our classically scale-invariant approach.

Ref.~\cite{Khoze:2013oga} used a  different realisation
of the leptogenesis mechanism proposed in \cite{Akhmedov:1998qx} and further developed in \cite{Asaka:2005pn,Drewes:2012ma} -- the leptogenesis
due to flavour oscillations of the right-handed Majorana neutrinos. It was shown in~\cite{Khoze:2013oga}  that  
the the classically-invariant 
SM$\times$U(1)$_{\sst B-L}$ model can generate the observed 
matter-anti-matter asymmetry on a large portion of the model parameter space without introducing any additional mass scales
and without
requiring any resonant fine-tuning.
The  Majorana masses are in the range
between 200 MeV and 500 GeV; they, together with and a heavier  mass for the Z' boson, are generated by the CW vev $\langle \phi \rangle$.
\bigskip

 (4.) {\it Vacuum stability:}
It is well-known that in the Standard Model the Higgs quartic coupling becomes negative rendering the Higgs effective potential 
unstable at above the scale $\simeq 10^{11}$ GeV. The next-to-next-to leading order analysis of the SM Higgs effective potential
carried out in \cite{Degrassi:2012ry} found the best-fit value of the Higgs self-coupling at the Planck scale to be small 
but negative,
\begin{equation}
  \label{cern}
  \frac{1}{2} \lambda_{\rm H}(M_{\rm Pl})\,=\, -0.0144 \,+ 0.0028\left(\frac{m_h}{\rm GeV} -125.66\right)\pm 0.0047_{M_t}
  \pm 0.0018_{\alpha_s} \pm 0.0028_{\rm th}\,,
\end{equation}
with further refinements carried out in \cite{Buttazzo:2013uya}.

It is remarkable that $\lambda_{\rm H}$ and its beta function nearly vanish at the Planck scale, but without new physics, 
the absolute stability of the Standard Model Higgs potential is ruled out \cite{Degrassi:2012ry,Buttazzo:2013uya}.

A minimal and robust way to repair the electroweak vacuum stability is provided by the Higgs portal extension
of the SM  -- and this is provided automatically in the classically scale-invariant theory \eqref{potentialcoupled1}.
There are two effects helping to stabilise the vacuum:
first is the fact that the portal coupling gives a positive contribution to the beta function of the Higgs quartic coupling,
$\Delta \beta_{\lambda_{\rm H}} \sim +\lambda_{\rm P}^2$. The second effect occurs due to a vev of the second scalar,
$\langle \phi \rangle > v$, leading to mixing between $\phi$ and the Higgs and resulting in a 
threshold correction lifting the SM Higgs self-coupling \cite{Lebedev:2012zw,EliasMiro:2012ay}. In the present context of 
classically scale-invariant SM $\times$ U(1)$_{\sst CW}$ theory, the vacuum stabilisation was addressed in recent papers  
Refs.~\cite{Iso:2012jn,Chun:2013soa,Hambye:2013dgv,Carone:2013wla}.
The values and ranges of the couplings needed to stabilise the Higgs potential depend on the model specific RG-running.\footnote{The authors 
of \cite{Hambye:2013dgv,Carone:2013wla} considered a model with the non-Abelian SU(2)$_{\sst CW}$ gauge group. 
Ref.~\cite{Hambye:2013dgv} showed that for the values of the portal coupling $\lambda_{p}=2 \lambda_{\rm P} \simeq 4 \times 10^{-3}$ and the non-Abelian
CW coupling $g_{\sst CW} \sim 1$, the vacuum stability is recovered (here the second scalar is heavy, $m_{h_2}\simeq 165$ GeV). 
More generally, in  Ref.\cite{Carone:2013wla}
the allowed regions on the entire parameter space $\{\lambda_{p},g_{\sst CW}\}$ consistent with the vacuum stability and perturbativity were 
determined for $m_{h_2} < m_{h_1}$ and for $m_{h_2} > m_{h_1}$, we refer to their Fig.~1 for more detail. 
Vacuum stability in the  SM $\times$ U(1)$_{\rm B-L}$ theory was addressed
in \cite{Chun:2013soa} and \cite{Iso:2012jn} following a more constrained than our approach, where the vanishing (rather than just positivity) 
of the Higgs potential was requested at $M_{\rm Pl}$.}  
The conclusion we draw from this work is that the vacuum stability up to the Planck scale is effectively restored on large portions of the parameter space for
classically conformally invariant extensions of the Standard Model. 

We also mention that adding an additional singlet to the theory coupled to the
Higgs via a portal will further enhance the stability of the potential.

\bigskip

 (5.-6.) {\it Inflation} and {\it dark matter} within the classically scale-invariant BSM formulation is the subject of the present paper.
 In the following section we will extend the Higgs portal model by introducing a single real scalar field $s(x)$
 with a non-minimal coupling to gravity. An exponentially flat potential for the singlet will be generated, as required for a successful
implementation of a slow-roll cosmological inflation, preserving the classical scale invariance of the model and without 
the need of perturbative unitarisation at sub-Planckian scales. 

Quite pleasingly, the singlet $s(x)$ is protected by a $Z_2$ symmetry and gives a viable candidate for dark matter. Thus the roles
of the inflation and of the scalar dark matter particle are unified and are straightforwardly incorporated in the context of the classically
scale-invariant Higgs-portal BSM model.

\bigskip
\section{Singlet-field slow-roll inflation in the Higgs portal}
\label{sec:2}

\medskip

In the standard Big Bang cosmology the cosmological inflation was proposed in~\cite{Guth:1980zm}
to solve the flatness, isotropy, homogeneity, horizon and relic problems. Inflation is the leading theory of the early universe,
confirmed by observations including the recent data from Planck satellite \cite{Ade:2013uln}
which favour a simple inflationary scenario with only one slow rolling scalar field.
 Slow-roll inflation occurs in many models constructed in the past,
although, in general, the underlying microscopic particle physics of many inflation models is still unknown, as it involves energy scales far higher than 
can be probed at colliders.

We will focus on the approach based on renormalisable QFT Lagrangians, 
which are then coupled to gravity, and in addition to the usual Einstein-Hilbert term 
also involve a non-minimal 
coupling of  a scalar field to gravity. By taking the non-minimal coupling $\xi$ to be large, a flat slow-roll potential is
generated and inflation takes place, as originally discussed in \cite{Salopek:1988qh}.

An interesting extremely minimal proposal which bridges cosmology with the Standard Model particle physics
is that the Higgs itself could play the role of the inflaton \cite{Bezrukov:2007ep}.
The coupling $\xi\sim 10^4$ however introduces two sub-Planckian scales in the theory,
$M_{\rm Pl} /\xi$ and $M_{\rm Pl} /\sqrt{\xi}$, where the theory appears to violate unitarity, thus spoiling the self-consistency 
of the inflationary approach, which has lead to a considerable debate in the literature as to whether such models make sense 
quantum mechanically~\cite{Barbon:2009ya,Burgess:2009ea,Lerner:2009na,Burgess:2010zq,Hertzberg:2010dc,Bezrukov:2010jz}.
From now and for the rest of this section we will be using the reduced Planck mass $ \sim 10^{18}$ GeV and will refer to it simply as $M$.

In \cite{Giudice:2010ka} it was shown that the Higgs inflation model can be unitarised by introducing (or integrating in) a new scalar
with the mass $M /\xi$ participating in the inflation. This allowed the authors of \cite{Giudice:2010ka} to raise the unitarity bound 
up to the Planck scale, well above the inflationary scale $\sim M /\sqrt{\xi}$. This approach however is incompatible with 
the requirement of classical scale invariance of the theory which we want to maintain. The mass $ M /\sqrt{\xi}$
of the new scalar field, which interacts with the Higgs, is a non-dynamical input scale. Upon integrating out this massive degree of freedom
the Higgs mass is destabilised and requires fine-tuning.

\medskip

We take an alternative approach to inflation which is based on a classically {\it massless} and vev-less real scalar field $s$. Our inflaton is neither the
Higgs itself as in Ref.~\cite{Bezrukov:2007ep}, nor the heavy scalar as in Ref.~\cite{Giudice:2010ka}. Our approach is also different from
the singlet-assisted Higgs portal inflation of Refs.~\cite{Lebedev:2011aq,Lebedev:2012zw} where the inflaton was a combination of the Higgs and the 
singlet with a non-vanishing vev (for earlier and related work see also \cite{Shaposhnikov:2006xi,Clark:2009dc,Lerner:2009xg,Bezrukov:2009yw,Lerner:2011ge}
and references therein). With the inflaton (or more precisely the non-minimally coupled to gravity scalar)
being a single-component field, we will be able to avoid problems with violations of unitarily at intermediate scales
in pure gravity and kinetic sectors, along the same lines
as in \cite{Lerner:2009na,Hertzberg:2010dc}.

\medskip

We start from the classical scalar potential \eqref{potentialcoupled1} of the SM coupled to the CW sector 
which we rewrite as
  \begin{equation}
    \label{potentialcoupled2}
    V_{\rm cl}(H,\phi)\,=\, \frac{\lambda_{\rm H}}{2}\left(|H|^2
      \,-\, \frac{\lambda_{\rm P}}{\lambda_{\rm H}}|\phi|^{2}\right)^2
    \,+\, \frac{\tilde\lambda_\phi}{4!}
    |\phi|^{4}\,.
  \end{equation}
  Here the new $\phi$-self-coupling  $\tilde\lambda_\phi$ is defined in terms of the original parameters of \eqref{potentialcoupled1} via a shift
 $\tilde\lambda_\phi = \lambda_\phi-12 \lambda_{\rm P}^2/\lambda_H$. 
 
 \medskip
 
We now extend this model by adding a single degree of freedom,  a one-component real scalar field $s(x)$. It is a gauge singlet 
which is coupled to the ${\rm SM}\times{\rm U(1)}_{\sst CW}$ 
theory
only via  the scalar portal interactions with the Higgs and $\phi$,
  \begin{equation}
    \label{potentialcoupled3}
    V_{\rm cl}(H,\phi,s)\,=\,  \frac{\lambda_{hs}}{2}|H|^2 s^2
    \,+\, \frac{\lambda_{\phi s}}{4} |\phi|^{2} s^2
    \,+\, \frac{\lambda_{s}}{4}  s^4 
     \,+\,   V_{\rm cl}(H,\phi)\,,
  \end{equation}
Equations \eqref{potentialcoupled2}-\eqref{potentialcoupled3} describe the general renormalisable gauge-invariant scalar potential 
for the three massless scalars,
\begin{equation}
  \label{rencond3}
  \frac{\partial^{2} V(H,\phi,s)}{\partial s \partial s}\bigg|_{s=H=\phi=0}=0\,,\quad
  \frac{\partial^{2} V(H,\phi,s)}{\partial H^{\dagger} \partial H}\bigg|_{s=H=\phi=0}=0\,,\quad
  \frac{\partial^{2} V(H,\phi,s)}{\partial \phi^{\dagger} \partial \phi}\bigg|_{s=H=\phi=0}=0\,.
\end{equation}
as required by the classical scale invariance of the theory.

The coupling constants in the potential \eqref{potentialcoupled3} are taken to be all positive (or at least non-negative), thus the potential is stable and
the positivity of $\lambda_{hs}$ and $\lambda_{\phi s}$ ensure that no vev is generated for the singlet $s(x)$. Instead the 
CW vev $\langle \phi \rangle$ generates the mass terms for the singlet,
\begin{equation}
    \label{ms2}
  m_s^2\,=\,  \frac{\lambda_{hs}}{2}\,v^2 
    \,+\, \frac{\lambda_{\phi ss}}{2} \, |\langle \phi \rangle|^{2}
    \,,
  \end{equation}
in the vacuum $s=0$, $\phi=\langle \phi \rangle$, $H=\frac{v}{\sqrt{2}}=\,\sqrt{ \frac{\lambda_{\rm P}}{\lambda_{\rm H}}}\, |\langle \phi \rangle |$.

We now proceed to couple our theory to gravity. To achieve the slow-roll inflation we need to introduce large non-minimal couplings to gravity of at least
some of the scalars\cite{Salopek:1988qh,Bezrukov:2007ep}. 
As noted earlier, the central point of our approach is to non-minimally couple  {\it only} the singlet $s$, such that the Lagrangian
takes the form,
\begin{eqnarray}
&&{\cal L}_{J} \,=\, \sqrt{-g_J} \left( -\, \frac{M^2}{2}R\,-\,\frac{\xi_s }{2}\, s^2 R  \,+\, \frac{1}{2} \,g_J^{\mu\nu}\,\partial_\mu s \,\partial_\nu s
\,+\, g_J^{\mu\nu}\,(D_\mu H)^\dagger D_\nu H
\,+\, \frac{1}{2} \,g_J^{\mu\nu}\,(D_\mu \phi)^\dagger D_\nu \phi
\right. \nonumber   \\
&& \left.- \, \frac{\lambda_{s}}{4}  s^4   \,-\, \frac{\lambda_{hs}}{2}|H|^2 s^2
    \,-\, \frac{\lambda_{\phi s}}{4} |\phi|^{2} s^2
     \,-\,   V(H,\phi) \,-\,\frac{1}{4}F^{\mu\nu}F_{\mu\nu}\,+\, {\rm Fermions} \,+\, {\rm Yukawas}\right)\,.
      \nonumber   \\
 \label{gravJ}
  \end{eqnarray}
The term $(\xi_s/2)\, s^2 R$ is the non-minimal coupling of the singlet $s$ to gravity, with 
$R$ being the scalar curvature. It will follow that for successful 
inflation the value of the non-minimal coupling constant $\xi_s$ should be relatively large, $\sim 10^4$.
For this reason, we will treat $\xi_s$ (and $\sqrt{\xi_s}$) as large parameters $\gg1$.
In this sense,  $s$ is distinguished from the two other scalars, $H$ and $\phi$,
which in our case have either vanishing or small loop-induced non-minimal gravitational couplings ($\xi_H \sim \xi_\phi\sim 1 \ll \xi_s$ and thus are neglected 
in \eqref{gravJ} without loss of generality). 

$M \sim 10^{18}$ GeV denotes the reduced Planck mass; in the Lagrangian \eqref{gravJ} $M$ appears only in the Einstein-Hilbert term
and does not couple directly to non-gravitational degrees of freedom. 
For the time being we will treat it as an explicit scale characterising the classical gravitational 
background.\footnote{To keep the scale invariance manifest in the full theory coupled to gravity, $M$ can be recast in terms of the vev of the dilation of the 
spontaneously broken local scale invariance. In dimensional regularisation, the scale-invariance can be kept manifest in quantum theory.
 We will return to this point in the Conclusions section.}
Finally, the subscript $J$ in \eqref{gravJ} defines the so-called Jordan frame, where the scalar-to-gravity coupling is non-minimal and kinetic terms are canonically normalised.

The second line on the {\it r.h.s.} of \eqref{gravJ} contains the scalar potential \eqref{potentialcoupled2} and \eqref{potentialcoupled3},
plus radiative corrections, plus kinetic terms for gauge and fermion fields and Yukawa interactions. 

An immediate question arises concerning the non-minimal coupling of the singlet $s$ to gravity, $\frac{\xi_s }{2}\, s^2 R$  in  \eqref{gravJ} :
Will the appearance of the relatively large coupling $\xi_s \sim 10^4$ render the theory ill-defined at an intermediate scale $M/\xi_s$ which is
below the high scale $M$? If one expands around the flat space, $g_{\mu\nu} = \eta_{\mu\nu} + h_{\mu\nu} /M$,
the non-minimal coupling generates the dimension-5
\begin{equation}
 \nonumber
 \sim \frac{\xi_s }{M}\, s^2\, \partial^2 \, h
\end{equation}   
interaction.
This interaction generates contributions to e.g. the $2 s \to 2s$ scattering processes mediated by the graviton exchange. However, it was shown in Ref.~\cite{Hertzberg:2010dc}
that when summed
over $s$, $t$  and $u$ channels, the leading order in $\xi_s$ contributions to this process cancel, and the first non-vanishing effects are suppressed by 
the high scale (Planck mass $M$); thus the  theory in this sector appears to be safe up to the gravity scale $M$, and as such is well-behaved at the intermediate
scale $M/\xi_s$.
For the cancellation to occur it is however essential that there is only a single scalar non-minimally coupled to gravity \cite{Hertzberg:2010dc}.
So, the theory coupled to gravity as defined by its Lagrangian \eqref{gravJ} in the Jordan frame appears to be safe up to the high UV scale $M$ and the
appearance of the relatively large non-minimal coupling  $\xi_s$ does not immediately invalidate the theory at lower intermediate scales. We will return to this point in what follows.

\medskip

We now perform a metric transformation to the Einstein frame where the non-minimal scalar to gravity interaction is removed,
\begin{equation}
\label{Einst}
g_{\mu \nu} \,\rightarrow\, \Omega^{-2}\, g_{\mu \nu}\, ,\qquad \Omega^2\,:= \, 1\,+\,\frac{\xi_s s^2}{M^2}\, ,
\end{equation} 
so that the Lagrangian becomes,
\begin{eqnarray}
&&{\cal L}_{E} \,=\,\sqrt{-g_E} \left( -\, \frac{1}{2}\, M^2 R  \,+\, 
\left(\frac{\Omega^2 + \frac{6\xi_s^2 s^2}{M^2}}{\Omega^4}\right)\frac{g_E^{\mu\nu}\,\partial_\mu s \,\partial_\nu s}{2}
\,+\, \frac{g_E^{\mu\nu}\,(D_\mu H)^\dagger D_\nu H}{\Omega^2}
\,+\, \frac{1}{2} \frac{g_E^{\mu\nu}\,(D_\mu \phi)^\dagger D_\nu \phi}{\Omega^2}
\right. \nonumber   \\
&& \left.- \, \frac{1}{\Omega^4}\left(\frac{\lambda_{s}}{4}  s^4   \,+\, \frac{\lambda_{hs}}{2}|H|^2 s^2
    \,+\, \frac{\lambda_{\phi s}}{4} |\phi|^{2} s^2
     \,+\,   V(H,\phi) \right)\,-\,\frac{1}{4}F^{\mu\nu}F_{\mu\nu}\,+\, \frac{\rm Fermions}{\Omega^3} \,+\, \frac{\rm Yukawas}{\Omega^4}\right)\,.
 \nonumber   \\    
 \label{gravE}
  \end{eqnarray}
As expected, the non-minimal interaction of $s$ with the scalar curvature $R$ has disappeared. 
The price for this is the non-canonical normalisation of the kinetic term for $s$. The factor 
$6\xi_s^2 s^2/{M^2}$ arises from the transformation of $R$,
 \begin{equation}
\label{Rtransf}
-\sqrt{-g_J}\, \Omega^2 R(g_J) \,=\, -\sqrt{-g_E}\,\Omega^2 R(g_E)\,-\, 6\Omega\,\partial_\mu(\sqrt{-g_E} \,g_E^{\mu\nu}\,
\partial_\nu \Omega)\,,
\end{equation} 
followed by the integration by parts in the second term. This indeed gives
\begin{equation}
\label{eqn:scale}
\frac{6\,\xi_s^2 \,s^2}{M^2} \times  \frac{g_E^{\mu\nu}}{2\, \Omega^2}
\,\,\partial_\mu s \,\partial_\nu s\,,
\end{equation} 
which appears in \eqref{gravE}. 

The expression \eqref{eqn:scale} describes a dimension-6 interaction suppressed by the scale $M/\xi_s$. 
This appears to be a dangerously low scale for the successful investigation of inflation which as we will see occurs at 
the larger scale $M/\sqrt{\xi_s}$. Of course this is the Einstein frame reflection of the same phenomenon we have already 
considered in the Jordan frame (see the paragraph above \eqref{Einst}).
For the same reason, in the class of models considered here, 
where the non-minimal coupling to gravity involves a {\it single} real scalar, the appearance of this scale in the kinetic term 
\eqref{eqn:scale} is harmless. It turns out that the contribution to $2 s \to 2s$ scattering of the 4-point vertex  \eqref{eqn:scale} is vanishing
when the external states are put on-shell  \cite{Hertzberg:2010dc}. This is true at tree-level and continues to hold to arbitrary 
loop order. The underlying reason for why the theory continues to be well-behaved at the intermediate scale (at least in the scalar-kinetic sector and in the 
scalar-gravity sector) is the existence of the field
redefinition in the Einstein frame which will render the kinetic term canonical. We will now proceed along these lines.

To bring the kinetic term for our singlet field $s$ into the canonical form we perform the field redefinition $s=s(\sigma)$
following \cite{Bezrukov:2007ep} and define the singlet field $\sigma(x)$ via
\begin{equation}
\label{eqn-sigma}
\sigma\,=\, {\rm sign}(s)\, \int_0^s ds \sqrt{\frac{1}{\Omega^2}+\frac{6\xi_s s^2}{M^2 \Omega^4}}
\,,
\end{equation} 
so that 
\begin{equation}
\label{eqn-sigmadiff}
\frac{d\sigma}{ds}\,=\,  \sqrt{\frac{1}{\Omega^2}+\frac{6\xi_s s^2}{M^2 \Omega^4}} \,=\,
\sqrt{\frac{\Omega^2 +\frac{3}{2}M^2(\partial_s \Omega^2)^2}
{\Omega^4}}
\,.
\end{equation} 
The kinetic term for the singlet in \eqref{Einst} is now brought to the canonical form,
\begin{equation}
\left(\frac{1}{\Omega^2}+\frac{6\xi_s s^2}{M^2 \Omega^4}\right)\frac{g_E^{\mu\nu}\,\partial_\mu s \,\partial_\nu s}{2}\,\,=\,
\frac{1}{2}\,g_E^{\mu\nu}\,\partial_\mu \sigma \,\partial_\nu \sigma\,.
\end{equation} 
We note that as pointed out in \cite{Hertzberg:2010dc}, this field redefinition is possible and works only for the case of a 
{\it single} real scalar field coupled non-minimally to gravity -- which is the case we are considering.\footnote{For multi-component 
scalars, such as for example the Higgs itself (4 real components) or the CW complex scalar $\phi$ (2 real components), 
 no field redefinition would have amounted to the
canonical normalisation of all field components. As a result, the perturbative unitarity would have been lost 
at the scale $M/\xi$ which is lower than the inflation scale $\sim M/\sqrt{\xi}$ we want to study. The use of the unitary gauge where 
only the radial component
remains and the extra degrees of freedom are gauged away only shifts the unitarity problem to the gauge-Higgs 
interactions and the longitudinal vector boson sector. }

At small field values, for example at the electroweak scale and all the way up to $s \lesssim 10^{14}$ GeV,
the field redefinition \eqref{eqn-sigma} amounts to 
\begin{equation}
\label{eqn-sigmaS}
\sigma(x)\,= \,s(x) \,
 , \quad {\rm for} \, \,\,
s\ll M/\xi_s
\,.
\end{equation} 
At higher values of $s$, it is easy to see that the solution for $\sigma$ is well approximated by 
the analytic expression \cite{Bezrukov:2013fka}
\begin{equation}
\label{eqn-sigmaL}
\sigma(x)\,=\, \sqrt{\frac{3}{2}}M\, \log \Omega^2 [s(x)]\ , \quad {\rm for} \, \,\,s \gg \frac{M}{\xi_s}
\,,
\end{equation} 
or
\begin{equation}
\label{eqn-sL}
s(x)\,=\, 
\frac{M}{\sqrt{\xi_s} }\,\sqrt{\exp\left(\frac{2\sigma(x)}{\sqrt{6}M}\right)-1}
\ , \quad {\rm for} \, \,\,s \gg \frac{M}{\xi_s}
\,,
\end{equation} 
At an even higher scale $s\gg M/\sqrt{\xi_s}$  this gives, 
\begin{equation}
\label{eqn-sigmaLL}
s(x) \,=\, \frac{M}{\sqrt{\xi_s} }\times \exp\left(\frac{\sigma(x)}{\sqrt{6}M}\right)\ , \quad {\rm for} \, \,\,s \gg \frac{M}{\sqrt{\xi_s}}
\,.
\end{equation}

From \eqref{eqn-sigmaL} and \eqref{eqn-sigmaL} we can compute 
\begin{equation}
\label{eqn-s2O2}
\frac{s^2(x)}{\Omega^2} \,=\, \frac{M^2}{\xi_s }\times \left(1-\exp\left[-\frac{2\sigma(x)}{\sqrt{6}M}\right]\right)\,\simeq\,\frac{M^2}{\xi_s }\, , \quad {\rm for} \,
 \,\,s \gg \frac{M}{\xi_s}
\,,
\end{equation}
and thus we find that at large field values, the singlet self-interaction potential 
in \eqref{gravE} is exponentially flat when expressed in terms of the canonically normalised $\sigma(x)$, and is well-suited for the slow-roll inflation
\cite{Bezrukov:2007ep},
\begin{equation}
\label{eqn-Vs}
V_E(s)\,:=\,\frac{\lambda_s}{4}\,\frac{s^4(x)}{\Omega^4} \,=\, \frac{\lambda_s M^4}{4\,\xi_s^2 } \left(1-\exp\left[-\frac{2\sigma(x)}{\sqrt{6}M}\right]\right)^2\, , \quad {\rm for} \,
 \,\,s \gg \frac{M}{\xi_s}
\,.
\end{equation}

\medskip

Next we need to take care of the inverse 
powers of  $\Omega$ in the Lagrangian \eqref{gravE}.
The appeared because the transformation \eqref{Einst} was a pure metric transformation, rather then the full Weyl scaling transformation,
\begin{equation}
\label{Weyltr}
{\rm Weyl:} \quad
g_{\mu \nu} \,\rightarrow\, \Omega^{-2}\, g_{\mu \nu}\, ,\quad 
{\rm scalars}  \,\rightarrow\, \Omega^1\, {\rm scalars} \, ,\quad 
A_\mu \,\rightarrow\, A_\mu \, ,\quad 
\Psi  \,\rightarrow\, \Omega^{3/2}\, \Psi\,.
\end{equation} 
We would now like to excise inverse factors of $\Omega$ from the Einstein-frame Lagrangian in \eqref{gravE}
to achieve the canonical normalisation for all the remaining kinetic terms. To do this, we now transform the scalar fields $H$ and $\phi$ (but not 
$s$ or $\sigma$!) as well as the fermions, as prescribed by the Weyl transformation  \eqref{Weyltr}. For the kinetic terms of non-singlet scalar fields
in \eqref{gravE}
(here we introduce the simplifying notation $\vec \varphi \,=\, (H/\sqrt{2}, \phi) \,= \, (h_1,h_2,h_3,h_4,\phi_1,\phi_2)$ for all six real components of the
Higgs and the CW scalar), with the transformation $\vec \varphi \,\rightarrow \,  \Omega\,\vec \varphi$  we have,
\begin{eqnarray}
 \frac{1}{2} \frac{g_E^{\mu\nu}\,\partial_\mu(\Omega\, \vec \varphi)\,\partial_\nu (\Omega\,\vec \varphi)}{\Omega^2} &=&
\frac{1}{2} {g_E^{\mu\nu}\,(\partial_\mu \vec \varphi)\,(\partial_\nu \vec \varphi)}
 \nonumber   \\
&+& 
\frac{1}{2} \frac{|\vec \varphi|^2}{6 M^2}\,g_E^{\mu\nu}\,(\partial_\mu \sigma)\,(\partial_\nu \sigma)\,+\,
\frac{\vec \varphi \cdot (\partial_\mu \vec \varphi)\,g_E^{\mu\nu}\, (\partial_\nu \vec \varphi)}{\sqrt{6}\, M}\nonumber   \\
&\simeq& 
\frac{1}{2} {g_E^{\mu\nu}\,(\partial_\mu \vec \varphi)\,(\partial_\nu \vec \varphi)}
\,,
 \label{kinphi}
  \end{eqnarray}
where we have used the substitution $\partial_\mu\, \log \Omega\,=\, \partial_\mu \sigma(x)/(\sqrt{6}M)$ valid at the singlet field values
$s \gg \frac{M}{\xi_s}$. The key point we want to make is that the two terms on the second line of \eqref{kinphi} are suppressed by the
Planck scale $\sqrt{6}\, M\, \gg \, M/\sqrt{\xi_s} \, \gg \, M/\xi_s$. The Planck scale is treated as the {\it large} scale $>$ than the UV cutoff in our effective field theory in the gravitational background, and we can neglect the terms suppressed by $M$. The inflation will happen at the parametrically  lower scale 
$\sim  M/\sqrt{\xi_s}$ dictated but the exponential fall-off of the potential \eqref{eqn-Vs}, and is under control in the theory with the canonically
normalised kinetic terms, given by the last line of \eqref{kinphi}.

The terms in \eqref{gravE} 
involving portal interactions or the singlet $s$ with other scalars, at large values of the singlet field (where 
$s^2/\Omega^2 = M^2/\xi_s$) become
\begin{equation}
\label{eqn-sPortal}
\frac{1}{\Omega^4}\left(\frac{\lambda_{hs}}{2}|\Omega H|^2 s^2
    \,+\, \frac{\lambda_{\phi s}}{4} |\Omega \phi|^{2} s^2 \right)\,\simeq\,
\frac{\lambda_{hs}M^2}{2\xi_s}   \,  |H|^2
\,+\, \frac{\lambda_{\phi s }M^2}{4\xi_s} | \phi|^{2} 
\,,
\end{equation}
giving the mass terms for the Higgs and the CW field in this large-$s$-field regime, 
with the singlet $s$ being (exponentially) decoupled from the rest of the theory, {\it c.f.}
Eq.~\eqref{eqn-s2O2}.
Finally,  the Higgs portal potential $V(H,\phi)$ is the homogeneous degree-4 expression in terms of scalars, given by the classical potential 
\eqref{potentialcoupled2} plus the radiative corrections also involving $\log (\phi/\langle \phi\rangle)$, such as the 1-loop CW expression,
(see the Appendix),
\begin{equation}
  \label{Veff2}
  V_{\sst CW}(\phi)\,=\,V_{\rm cl}
  \,+\,
  \frac{3}{4} \alpha_{\sst CW}^{2} |\phi|^{4}
  \left[\log\left(\frac{|\phi|^{2}}{\langle |\phi|^{2}\rangle}\right)
    -\frac{25}{6}\right]\,.
\end{equation}
Since $V$ goes as a sclar field to the 4th power,
the factors of $\Omega$ cancel,
\begin{equation}
\label{eqn-HPortal}
\frac{1}{\Omega^4}V(\Omega H,\Omega \phi) \,=\, V( H, \phi)
\,.
\end{equation}

To summarise, the Einstein-frame Lagrangian after the Weyl transformation of all the fields, except for 
the singlet $s(x)$ which is substituted by the field redefinition \eqref{eqn-sigma}, at large values of the singlet field takes the canonical form. The
 singlet in this regime is effectively decoupled from the SM degrees of freedom and the exponentially flat potential \eqref{eqn-Vs},
 \begin{eqnarray}
&&\frac{{\cal L}_{E}}{\sqrt{-g_E}} \,=\, -\, \frac{1}{2}\, M^2 R  \,+\, 
\frac{1}{2}\,\partial^\mu \sigma \,\partial_\mu \sigma
\,- \, \frac{\lambda_s M^4}{4\,\xi_s^2 }  \left(1-\exp\left[-\frac{2\sigma}{\sqrt{6}M}\right]\right)^2 \,-\,\frac{1}{4}F^{\mu\nu}F_{\mu\nu}
\,+\, {\rm Fermions}
\nonumber   \\
&& +\, (D^\mu H)^\dagger D_\mu H
\,+\, \frac{1}{2} \,(D^\mu \phi)^\dagger D_\mu \phi  \,-\, 
\frac{\lambda_{hs}M^2}{2\xi_s}   \,  |H|^2
\,-\, \frac{\lambda_{\phi s} M^2}{4\xi_s} | \phi|^{2} 
     \,-\,   V(H,\phi)  \,+\, {\rm Yukawas}\,.
 \nonumber   \\    
 \label{gravW}
  \end{eqnarray}
In the equation above we have used the exponential form potential \eqref{eqn-Vs} for the canonically normalised $\sigma$-singlet 
which is the correct description at large field values. At the same time, to simplify the formulae, in the terms involving portal interactions between the 
singlet and other scalars, we used the asymptotic values for $s^2/\Omega^2$ as in \eqref{eqn-sPortal}, emphasising the emerging large mass terms 
for $H$ and $\phi$, but neglecting exponentially suppressed corrections involving $\sigma$. These are small exponentially suppressed effects
which are not going to modify the analysis of inflation.

The inflaton is the $\sigma$-field and the inflation is generated by the
 self-interacting part of the potential in \eqref{gravW}, 
\begin{equation}
\label{eqn-Vsigma}
V(\sigma)\,=\,\ \frac{\lambda_s M^4}{4\, \xi_s^2 } \left(1-\exp\left[-\frac{2\sigma(x)}{\sqrt{6}M}\right]\right)^2\, .
\end{equation}
It is exponentially flat for large field values and is suitable for the slow-roll inflation \cite{Bezrukov:2007ep}.
The slow-roll inflation parameter is
\begin{equation}
\label{eqn-eps}
\epsilon\,:=\, \frac{M^2}{2}\left(\frac{V(\sigma)/d\sigma}{V(\sigma)}\right)^2\,=\, \frac{4\,M^4}{3\,\xi_s^2 \,s^2}\, .
\end{equation}
Inflation ends when $\epsilon_{\rm end}=1$ which corresponds to
$s_{\rm end} = (4/3)^{1/4} M/\sqrt{\xi_s}$ or $\sigma_{\rm end} \simeq 0.94 M$.
Inflation starts at the singlet field value  \cite{Bezrukov:2013fka} $s_{0} \simeq 9.14 \, M/\sqrt{\xi_s}$.
The CMB normalisation condition,
\begin{equation}
\label{eqn-CMB}
\frac{V}{\epsilon} (s=s_o)\,\simeq\, (0.0276\, M)^4\, ,
\end{equation}
which determines the non-minimal singlet coupling to gravity \cite{Bezrukov:2007ep,Bezrukov:2013fka},
\begin{equation}
\label{eqn-xi-s}
\xi_s\,\simeq\, 4.7 \times 10^4 \, \sqrt{\lambda_s}\, .
\end{equation}

\bigskip

The spectral index and the tensor-to-scalar perturbation ratios in this model are  the same as computed in the Bezrukov-Shaposhnikov 
Higgs-inflation model \cite{Bezrukov:2007ep}. They are found to be in excellent agreement  \cite{Bezrukov:2013fka}
with the latest Planck measurements, \cite{Ade:2013uln}. 

\bigskip

Our realisation of inflation in \eqref{gravW} is a one-field slow-roll 
inflation model. The singlet field $\sigma$ plays the role of the inflaton, while the other degrees of freedom,
such as the Higgs and the CW scalar at the inflation scale being much heavier than the Hubble expansion parameter during inflation,
$H = \, \sqrt{\frac{\lambda_s}{12}}\, \frac{M}{\xi_s}.$
\begin{equation}
\label{eqn-decoupl}
m_{h} \, =\, \sqrt{ \frac{\lambda_{hs}}{2}}\, \frac{M}{\sqrt{\xi_s}}
\quad {\rm and}\quad 
m_{\phi} \,=\,  \sqrt{ \frac{\lambda_{\phi s}}{2}}\, \frac{M}{\sqrt{\xi_s}}
\quad \gg\quad
H\, = \, \sqrt{\frac{\lambda_s}{12}}\, \frac{M}{\xi_s}\,.
\end{equation}

\bigskip

The main technical advantage of the singlet field realisation of inflation presented here, compared to the 
original Higgs inflation model \cite{Bezrukov:2007ep} and the subsequent implementations with other multi-component non-minimally coupled scalars, as in 
\cite{Lebedev:2011aq,Lebedev:2012zw}, is that our model does not require introduction of new physics effects associated with the `low' $ M/\xi_s$ and 
`intermediate' scale $ M/\sqrt{\xi_s}$. Indeed, the appearance of the low-scale dimension-6 operator in \eqref{eqn:scale} was rendered
harmless by the field redefinition of the one-component field $s(x)$.

At the same time we should note that the field redefinition \eqref{eqn-sigma}-\eqref{eqn-sigmadiff} when applied in the regime of 
{\it small} field values, $s \ll M/\xi_s$, does introduce higher-dimensional corrections in the Einstein-frame potential for the singlet, {\it cf} \eqref{eqn-Vs},
\begin{equation}
V_E(s)\,=\,\frac{\lambda_s}{4}\,\frac{s^4(x)}{\Omega^4} \,=\,\frac{\lambda_\sigma}{4}\,s^4(x) \,-\, 
\frac{\lambda_s}{4}\, \frac{\xi_s^2}{M^2}\,\sigma^6(x) \,+\, \ldots
\, , \qquad {\rm for} \,
 \,\,s \ll \frac{M}{\xi_s}
\,.
\end{equation}
If one considers multi-particle (2-to-4 and higher) scattering processes,
this appears to break unitarity at scales $\frac{M}{\xi_s}$ in the perturbative regime of the theory when expanding at small values of $\sigma$.  
These scattering processes are relevant for the theory in the `collider regime' i.e. in the regime where we are using perturbation
theory around vanishing background field values. However in the collider regime the theory is safe:  scales $\sim \frac{M}{\xi_s}$ 
are of course inaccessibly high for any collider experiments.
On the other hand, in order to probe the theory around the scale of inflation, 
the more appropriate approach is to expand around a non-trivial background. In this case,
 as argued in Ref.~\cite{Bezrukov:2010jz}, the effective UV-cutoff scale of the theory itself depends on the background field values (i.e. in our case 
 it depends on $\bar s$ or $\bar \sigma$) and rises with these values appropriately. The (slightly simplified) results of the analysis in \cite{Bezrukov:2010jz}
 imply for the effective UV cut-off of theory, $\Lambda_{\rm eff} (\bar s) \sim  \frac{M}{\xi_s}$ for $\bar s \ll \frac{M}{\xi_s}$;
$\Lambda_{\rm eff} (\bar s) \sim  \frac{\xi_s\, {\bar s}^2}{M}$ for $\frac{M}{\xi_s} \ll \bar s \ll \frac{M}{\sqrt{\xi_s}}$, and finally, 
$\Lambda_{\rm eff} (\bar s) \sim  M$ for $ \bar s \gg \frac{M}{\sqrt{\xi_s}}$ in the inflationary regime. We conclude that the effective potential 
of the theory is likely to remain safe
at all relevant energy scales (below the Planck scale) when evaluated at the appropriate background field value 
(see Ref.~\cite{Bezrukov:2010jz} and references therein for the detailed analysis of quantum corrections).

\medskip 

We also note that the fact that all the terms for the non-singlet scalars, $H$ and $\phi$, are already canonically normalised in the
Lagrangian 
\eqref{gravW} implies that there are no non-renormalisable interactions in the theory involving the sub-Planckian scales.
Indeed, the only terms we dropped in \eqref{kinphi} were suppressed by the high scale $M$. This is different from the case of the Higgs-driven inflation  
where the gauge and fermion interactions of the Higgs are modified and have to be cut-off at the intermediate scale $ M/\sqrt{\xi_s}$, as pointed out in
\cite{Giudice:2010ka}.

\bigskip

It is important to note that if the new physics was required to be included at the scales below the UV cutoff $M$,
this would have destroyed the classical scale invariance of our model and induced large threshold contributions to the
masses of the Higgs and the CW scalar, reintroducing the fine-tinning problem into the SM. For this reason the elegant 
construction of Ref.~\cite{Giudice:2010ka} which integrates in a massive linear sigma-model field at the scale $ M/\xi_s$ to restore the
perturbative unitarity of the Higgs inflation model does not work in our classically scale-invariant case.

\bigskip

Finally, we would like to briefly comment on the role of loop-induced quantum corrections to the effective potential.
As explained in the Introduction, our calculational approach is based on the use of dimensional regularisation where no power-like divergencies can appear.
 At the 1-loop level 
we have the Coleman-Weinberg corrected potential,
\begin{equation}
V_{\rm 1-loop} (\sigma)\,=\, V_{\rm cl } (\sigma)\,+ \,
\frac{1}{64 \pi^2} \left(V_{\rm cl}''(\sigma)\right)^2\,\log\left(V_{\rm cl}''(\sigma)/M_{\sst UV}^2\right)^2\,,
\end{equation}
which entails the derivatives of the exponentially slowly varying potential in the background of  a large $\sigma$-field.
Higher loop orders will involve higher derivatives of the flat potential enhanced by only a logarithmic dependence on the
UV-cutoff. 
As a result, it is quite likely that  quantum corrections computed in the dimensional regularisation scheme 
in the background perturbation theory around a large $\sigma$-field will be small and not affect the picture of inflation
at the leading order, see also \cite{Bezrukov:2010jz} for a more comprehensive discussion of this point.

\bigskip
\section{Dark matter}
\label{sec:3}    

Singlet scalar dark matter (DM) models \cite{Silveira:1985rk}
are the simplest possible UV-complete models of dark matter containing a weakly interacting massive
particle (WIMP) -- in this case a scalar interacting with the Higgs via the portal interaction.

In the classically scale-invariant SM$\times$U(1)$_{\sst CW}$ theory with the real singlet 
$s(x)$ and the scalar potential \eqref{potentialcoupled3}, the stability of the singlet is protected by 
a $Z_2$ symmetry, $s \to -s,$ giving a natural dark matter candidate. Indeed, the $Z_2$ symmetry
of the potential \eqref{potentialcoupled3}
is an automatic consequence of the renormalisability (dimension 4), scale-invariance and gauge invariance 
(which does not allow odd powers of $H$ and $\phi$) of the theory. 

As shown in the previous section, at large field values, $s(x) > M/\sqrt{\xi_s}$, the singlet (or more precisely its log in
\eqref{eqn-sigmaLL}) 
plays the role of the inflaton field which during the inflation slowly rolls in an exponentially flat potential \eqref{eqn-Vsigma}.
In this regime, the inflaton gives large masses \eqref{eqn-decoupl} to the SM fields and otherwise decouples from the
rest of the theory. 

After inflation is completed, the singlet enters the regime $s(x) \ll M/\xi_s$ where it is canonically normalised, its
potential is no longer flat and given by  \eqref{potentialcoupled3}, and the large masses of the SM fields are no longer there. 
The singlet now assumes the role of the dark matter in the classically scale-invariant theory.

\medskip

There are only two phenomenologically relevant parameters of this singlet dark matter model,
the coupling $\lambda_{hs}$ of the dark matter particle to the SM Higgs,
and the induced mass of the singlet, $m_s \equiv m_{\sst DM}$ in Eq.~\eqref{ms2} which gives the dark matter mass.
Expressed in terms of these two parameters, the DM phenomenology of this classically scale-invariant model is
the same as that of the usual singlet scalar DM model.

\begin{figure}[t!]
\begin{center}
\begin{tabular}{cc}
\hspace{-.4cm}
\includegraphics[width=0.5\textwidth]{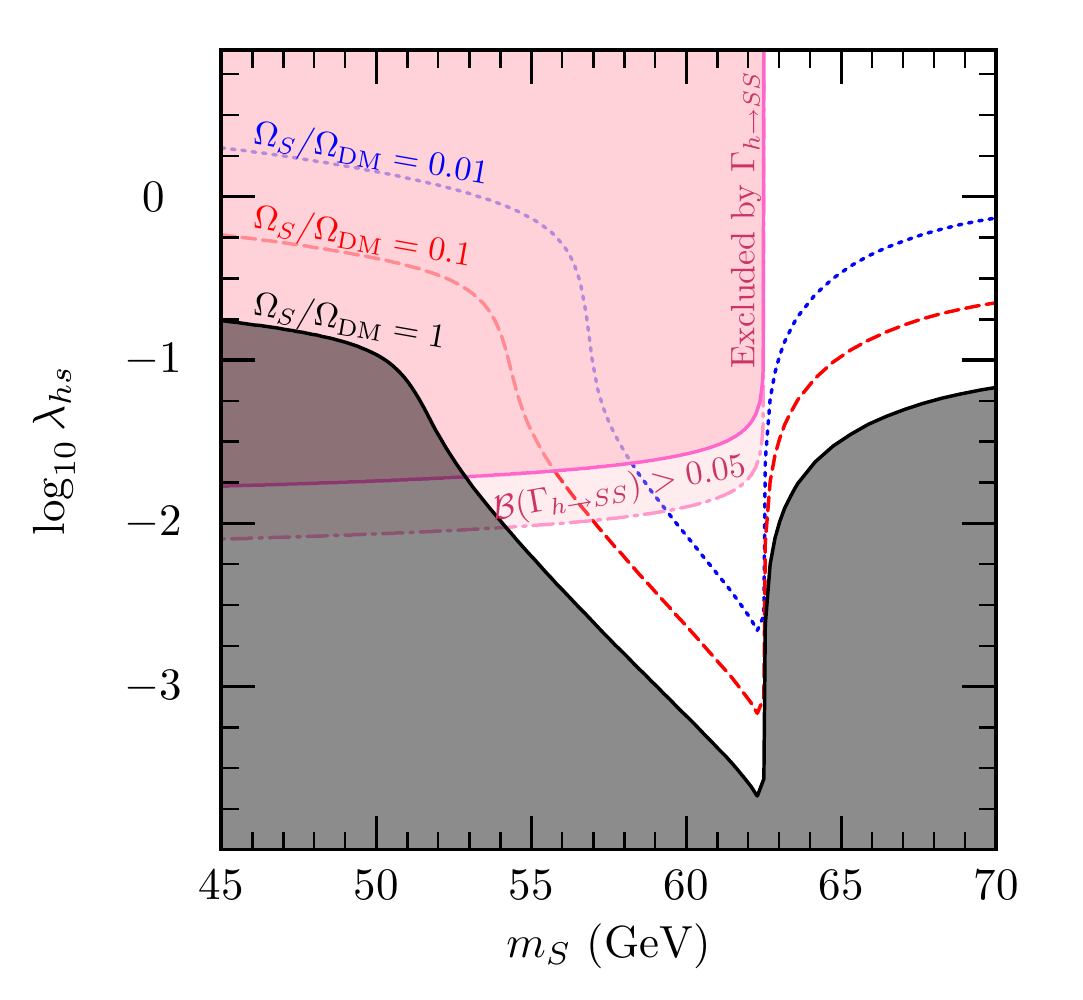}
&
\includegraphics[width=0.5\textwidth]{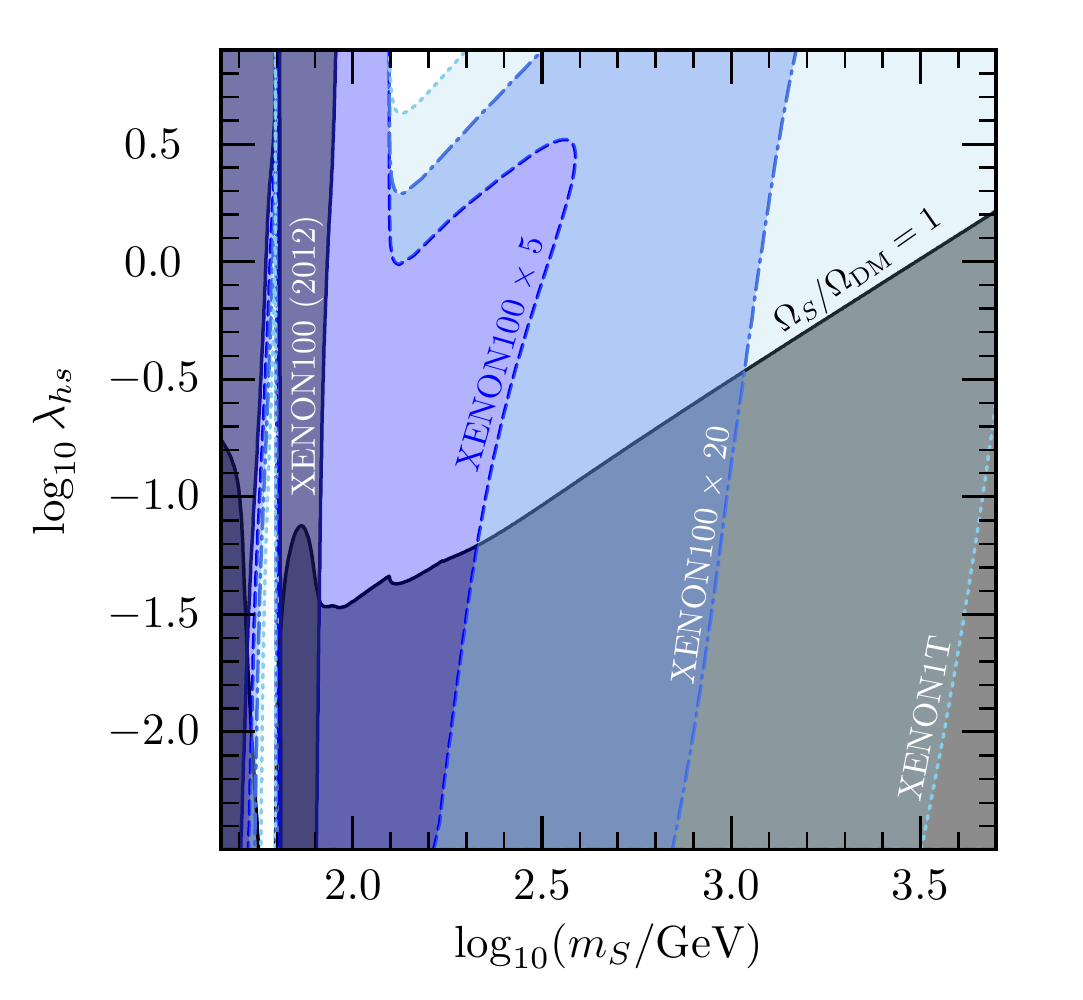}
\\
\end{tabular}
\end{center}
\vskip-.4cm
\caption{
\label{fig:DM}
\small {\bf DM exclusion contours from Ref.~\cite{Cline:2013gha}} on the $\lambda_{hs}$, $m_s$ plane. 
\textit{Left} plot is a close-up on the region $m_s \lesssim m_h/2$. 
The dark-shaded lower region is ruled out by the upper bound on the singlet DM relic density, $\Omega_S/\Omega_{\rm DM} \le 1$.
The region in the upper-left corner is ruled out by constraints on invisible Higgs decays.
\textit{Right} plot is over a wide mass range $45\,{\rm\,GeV} \le m_s \le 5\,{\rm TeV}$. The DM relic density exclusion is shown superimposed with the current bound from XENON100 and bounds from future direct detection experiments (in lighter shades of blue).}
\end{figure}

Papers \cite{Mambrini:2011ik,Englert:2011aa,Djouadi:2011aa,Low:2011kp,Cheung:2012xb,Cline:2013gha}
contain the recent comprehensive studies of the singlet scalar DM, see also references therein. Here we give only a very brief summary
of their results. The allowed/excluded portions of the $\lambda_{hs}$, $m_s$ parameter space are shown on Figure~\ref{fig:DM}
taken from Ref.~\cite{Cline:2013gha}.

1. The relic density of singlet scalar dark matter, $\Omega_S$, should not exceed the full DM relic density $\Omega_{\rm DM}.$
The DM relic density is determined by the (in-)efficiency of the DM particles annihilating into the SM degrees of freedom. These processes are
dominated by the Higgs-mediated s-channel. Essentially the constraint $\Omega_S/ \Omega_{\rm DM}\,\le 1$ amounts
to a lower bound on the coupling $\lambda_{hs}$. The excluded region is the lower portion of the parameter space shown in
dark grey on both plots of Fig.~\ref{fig:DM}.

2. Indirect detection of dark matter constraints arise from annihilation of DM to SM particles. For singlet scalar DM these constraints work
much in the same way as the relic density constraint above,  and for $m_s> m_{h}/2$ provide a slight improvement on the relic density contours.

3. When $m_s <  \frac{1}{2}\,m_h$ the DM can be produced via invisible Higgs decays at the LHC
with the Higgs partial decay widths into the singlet DM particles given by
\begin{equation}
\Gamma_{h\to ss}\,=\, \frac{\lambda_{hs}^2\,v^2\,\sqrt{1-4m_s^2/m_h^2}}{16 \pi\,m_h}\,.
\end{equation}
The upper bounds on the invisible Higgs decays from the LHC give the exclusion contour 
shown in red on the upper-left corner of the left plot of Fig.~\ref{fig:DM} which zooms into
the singlet mass region between 45 and 70 GeV. Not surprisingly, the bounds from invisible Higgs decays and 
from the DM relic density work in the opposite way, with the former excluding larger values of the coupling $\lambda_{hs}$
in the kinematically allowed phase space. 

4. Direct detection experiments
provide upper bounds on the DM-nucleon cross-sections.
Here DM interacts elastically 
with nuclei through the Higgs boson exchange. The resulting nuclear recoil is then interpreted in terms of the
DM mass and DM-nucleon cross section.
The spin-independent cross-section is
\begin{equation}
\sigma^{\rm SI}\,=\,
\frac{\lambda_{hs}^2}{4 \pi\,m_h^4}\,\frac{m_N^4 f_N^2}{(m_s+m_N)^2}
\,,
\end{equation}
where $f_N$ parameterises the Higgs-nucleon coupling.
These bounds are translated into exclusion contours on the ($\lambda_{hs}$, $m_s$) plane, which essentially are lower bounds on
the scalar DM mass $m_s$. Constraints from the current XENON100 direct detection experiment \cite{Aprile:2012nq}
are shown in dark blue on the left of second plot in Fig.~\ref{fig:DM} together with the future reach of XENON upgrades 
in lighter shades of blue, as computed in Ref.~\cite{Cline:2013gha}.

\medskip

To summarise, the singlet scalar DM model is perfectly viable at present on a large portion of its parameter space
for $m_s > \frac{1}{2}\,m_h$. At lower values of $m_h$ in the region between 55 GeV and $\frac{1}{2}\,m_h$, there
is only a small triangle (in white) left unexplored, as can be seen from the plot on the left.
More generally, the XENON100 upgrade and the XENON1T direct detection experiment should be able to probe 
the entire parameter space of the singlet DM model \cite{Djouadi:2011aa,Cline:2013gha}.

\section{Conclusions}
\label{sec:five}

In this paper we considered a classically scale-invariant SM theory extended with the Coleman-Weinberg sector and an additional real singlet field,
both coupled to the SM via the Higgs portal interactions. The singlet classically massless field $s(x)$ was shown to give
rise to the inflaton and to the dark matter candidate. 

We have presented for the first time an implementation of the slow-roll inflation mechanism 
in a  BSM theory with classical scale invariance. Our model is a single-real-field inflationary model
which does not suffer from a breakdown of unitarity at scales below or comparable to the scale of inflation.
The model maintains its classical scale invariance and is self-consistent all the way up to the Planck-scale UV cutoff.
The SM Higgs potential is stabilised by the Higgs portal interactions with the Coleman-Weinberg scalar.
The Higgs mass and the scale of the electroweak symmetry breaking are unaffected by 
the inflaton and its interactions.
Furthermore, the singlet field responsible for inflation also gives a viable scalar dark matter candidate in our model.

\bigskip 

When discussing cosmological inflation in Section 2, our classically scale invariant quantum field theory was coupled 
to a gravitational background with an explicit dependence on the reduced Planck scale $M$. How does this fit with 
classical scale invariance of the non-gravitational theory? 

The Planck scale $M$, as it appears in our effective Lagrangians \eqref{gravJ}
and \eqref{gravW} does not cause problems to the effective field theory. At the scale $M$ and above, the theory
coupled to gravity of course becomes non-renormalisable, but $M$ is above the UV cutoff scale of our theory
and gravity is treated as an external background. One can continue using dimensional regularisation in the
curved background for all QFT processes.
The interesting for us phenomena, such as the cosmological inflation, happen at large values of the singlet field
set by the intermediate scale $M/\sqrt{\xi_s} \ll M$, at this scale the theory is robust, the scalar inflaton $\sigma(x)$ is exponentially 
decoupled from the SM degrees of freedom (in the relevant for inflation regime of large field values) and there are no heavy degrees 
of freedom coupled to the SM sector present in the theory.

Still it is desirable to not have an explicit scale $M$ present in the theory which is supposed to become classically scale-less
when gravity decouples.
One can do this and embed the classical scale-invariance of the theory in the IR into 
the scale invariance as a fundamental symmetry of nature - including gravity - with the Planck scale being set by the vev of 
the dilaton.
We stress first that this is not a necessary requirement,
the classical scale invariance of the IR theory can in principle exist even if the UV-complete theory is not scale-invariant.
Secondly, as we will explain below,  the classical scale invariance of the effective QFT does not automatically follow 
from the (broken) scale invariance of the gravitational theory. It has to be seen as an additional constraint.

To proceed, we impose the scale-invariance on the full theory including gravity and assume that this 
is a {\it local} symmetry which is non-linearly realised --
i.e. spontaneously broken by the dilaton field with the vev $\langle \varphi_{\rm dil}\rangle = M$.
Very recently there has been 
a resurgence of interest in such an approach to gravity and cosmology, 
see Refs.~\cite{Kallosh:2013pby,Bars:2013yba} and  also
\cite{Englert:1976ep,Shaposhnikov:2008xi,GarciaBellido:2011de} and references to earlier work therein for 
other related approaches which assume or utilise full quantum scale invariance of the underlying theory.

Thus our original Lagrangian \eqref{gravJ} can be 
written in a manifestly locally scale-invariant (Weyl-invariant) form if we retrofit the Planck scale in terms of the dilaton vev,
$\langle \varphi_{\rm dil}\rangle = M$ and introduce the kinetic term for the dilaton (see section III of \cite{Bars:2013yba}
for a review of a general construction),
\begin{equation}
 {\cal L}_{J} \,=\, \sqrt{-g_J} \left( -\, \frac{\varphi_{\rm dil}^2 +\xi_s s^2}{2}\, R  \,-\,
  \frac{1}{2} \,g_J^{\mu\nu}\,\partial_\mu  \varphi_{\rm dil}\,\partial_\nu \varphi_{\rm dil}
 \,+\, \frac{1}{2} \,g_J^{\mu\nu}\,\partial_\mu s \,\partial_\nu s 
\,+\, \ldots \,+\, V \right)\,,
   \label{gravWloc}
  \end{equation}
where the dots indicate omitted kinetic terms of other fields, and $V$ is the scalar potential which must scale as the 4th power of the 
scalars, which our potential does (see the discussion above \eqref{Veff2}).
The dilaton can be expressed as,
\begin{equation}
\varphi_{\rm dil} (x) \,=\, M \times \exp\left(\frac{\chi(x)}{M}\right)\,, 
 \label{eqn:compensator}
  \end{equation}
where $\chi$ transforms linearly under local scale transformations.
Note that the dilaton kinetic term in \eqref{gravWloc} has the `wrong' sign, but this is not a problem \cite{Kallosh:2013pby,Bars:2013yba},
since the dilaton is not a physical degree of freedom - it can be gauged away. (This also means that there are no additional massless
or light degrees of freedom arising from the dilaton, thus no complications for the SM.) In the unitary gauge we have
$\varphi_{\rm dil} = M.$

The dilaton spontaneously breaks local scale invariance of the gravitational theory at the scale $M$, and gives the mass gap $\sim M$
to heavy gravitational states.
In order to maintain the classical scale invariance of the SM theory, the graviton should not couple to any of the SM states in the
potential $V$ in \eqref{gravWloc}. This should be seen as an additional requirement (rather than an automatic consequence) 
to the local scale invariance of the microscopic theory. 

When Weyl transforming to the Einstein frame, as in section 2, we now promote $M$ to the dilaton field and maintain the
manifest local scale invariance, or gauge fix to the unitary gauge and keep $M$ as in section 2.
Quantum effects are computed in dimensional regularisation in the manifestly scale-invariant way. 

In the flat background and without scale invariance, the dimensionless coupling constants acquire the dependence on the RG scale $\mu$ in $D=4+\epsilon$
dimensions, $ \lambda \to \lambda \, \mu^\epsilon$. In the gravitational background,
one treats the metric tensor 
$g_{\mu\nu}$ as a $(4+\epsilon) \times (4+\epsilon)$ matrix, and in the Einstein frame, it introduces the dependence on the dilaton. 
The result is that in our theory with spontaneously broken (or non-linearly realised scale-invariance), the every appearance of the RG
scale $\mu$ in the effective potential is now substituted by, see Ref.~\cite{Sundrum:2003yt},
\begin{equation}
\mu \,\to\, \mu \times \exp\left(\frac{\chi(x)}{M}\right)\,,
 \label{eqn:compensator-mu}
  \end{equation}
  which now transforms under scale-transformations.
This gives a manifestly scale-invariant formulation in quantum theory regularised in dimensional regularisation.

\bigskip

The message we take from the above discussion is that the classical scale invariance of the effective SM theory 
can be compatible with, and form a natural part of the full local scale invariance of the full theory.
This, however, is not automatic and requires an additional constraint that the dilaton vev $\langle \varphi_{\rm dil} \rangle = M$ 
does not appear in the SM potential, though $\varphi_{\rm dil}/M$ can appear in the logarithms.
After the gravity is decoupled, together with the heavy states whose masses are set by the dilaton mss gap,
the SM $\times$ CW $\times$ $s$  theory  in the IR is classically scale invariant. This classical scale invariance 
has its own pseudo-dilaton -- the CW field $\phi$, which is a physical degree of freedom and has nothing to do with the 
heavy dilaton of the gravitational theory.

\bigskip

\section*{Acknowledgements}
We acknowledge the hospitality of the Aspen Center for Physics and thank participants for useful discussions. 
We also thank Christoph Englert, Joerg Jaeckel, Gunnar Ro and Michael Spannowsky for exchange of ideas and collaboration on related topics.
This material is based upon work supported in part by 
STFC through the IPPP grant ST/G000905/1, in part by the Wolfson Foundation and Royal Society and in part by the National Science Foundation Grant No. PHY-1066293.

\bigskip

\newpage
\startappendix
\Appendix{Mass generation in a classically massless theory}
\label{sec:app}

The simplest theory where the CW mechanism~\cite{Coleman:1973jx} is realised is the massless scalar QED
with the classical potential 
\begin{equation}
  V_{\rm cl}=\frac{\lambda_{\phi}}{4!}|\phi|^4\,.
\end{equation}
The field $\phi$ is the Coleman-Weinberg complex scalar, it is charged under the U(1)$_{\sst CW}$ gauge
symmetry with the usual covariant derivative coupling,  $D_\mu \phi = \partial_\mu \phi + i g_{\sst CW} \, A_{\mu}\, \phi$.

Normalisation conditions for $\phi$ 
are chosen to be the same as in \cite{Coleman:1973jx},
$\phi=\phi_{1}+i\phi_{2}$, with canonical kinetic terms for its real and imaginary part components
    $\frac{1}{2}(\partial_{\mu}\phi_{1}\partial^{\mu}\phi_{1}+
    \partial_{\mu}\phi_{2}\partial^{\mu}\phi_{2})$. This is related to the canonically normalised 
    complex scalar $\Phi=(\phi_{1}+i\phi_{2})/\sqrt{2}$ via a simple rescaling,
    $\Phi=\phi/\sqrt{2}$.

The 1-loop corrected scalar potential is given by the usual Coleman-Weinberg expression~\cite{Coleman:1973jx},
\begin{equation}
  \label{Veff1}
  V(\phi)\,=\, V_{\rm cl}+\Delta V_{\rm 1-loop}\,=
  \, \frac{\lambda_{\phi}}{4!}|\phi|^{4}\,+\,
  \left(\frac{5\lambda_{\phi}^2}{1152\pi^2}
    +\frac{3 g_{\sst CW}^{4}}{64\pi^2}\right)|\phi|^{4}
  \left[\log\left(\frac{|\phi|^{2}}{\mu^{2}}\right)
    -\frac{25}{6}\right]\,,
\end{equation}
where $\mu$ is the RG scale.
The effective potential above is computed in the massless theory,
in the UV subtraction scheme is 
chosen to set the renormalised mass at the origin of the field space to zero,
\begin{equation}
  \label{eq:m00}
  m^2(\phi=0;\mu) \,= \, V^{\prime\prime}(\phi)\bigg|_{\phi=0}=0 \, .
\end{equation}
We use dimensional regularisation which preserves classical scale-invariance of the theory. In dimensional 
regularisation the equation \eqref{eq:m00} is satisfied automatically and is independent of the RG scale $\mu$.

The minimum of the Coleman-Weinberg effective potential occurs at $\phi=\langle \phi \rangle$
where the first derivative of \eqref{Veff1} vanishes. Validity of weakly-coupled perturbative approach 
requires that $\lambda_{\phi} \sim \alpha_{\sst CW}^{2}$ in the minimum (rather than the more usual 
relation $\lambda_{\phi} \sim \alpha_{\sst CW}^{1}$) where we defined $\alpha_{\sst CW}=g^2_{\sst CW}/(4\pi)$.
This is because the minimum of the effective potential \eqref{Veff1} is determined  \cite{Coleman:1973jx} by ballancing the tree-level 
contribution $\sim \lambda_{\phi}$ against the 1-loop term $\sim \alpha_{\sst CW}^{2}$. 
Thus one can neglect in the $\lambda_{\phi}^2\sim \alpha_{\sst CW}^{4}$ 1-loop contribution on the right hand side of \eqref{Veff1} which is subleading 
relative to the $\alpha_{\sst CW}^{2}$  term.
Choosing the value of the RG scale $\mu=\langle \phi \rangle$,
the minimum of the 1-loop corrected effective potential occurs  at \cite{Coleman:1973jx}
\begin{equation}
  \label{Veffprime}
  \partial_\phi V\,=\, \left( 
  \frac{\lambda_{\phi}}{6}-11\,\alpha_{\sst CW}^{2}
  \right)
  {\langle \phi\rangle}^3 \, =\, 0
\end{equation}
and the vev $\langle \phi\rangle$ is determined by the condition on
the couplings renormalised at the scale of the
vev,
\begin{equation}
  \label{eq:rad4}
  \lambda_{\phi}(\langle\phi\rangle) \,=\, 66 \,\alpha_{\sst CW}^{2}(\langle \phi\rangle) 
  \,\equiv\, \frac{33}{8\pi^2} \,g^4_{\sst CW}(\langle \phi\rangle)\,.
\end{equation}
An RG point can always be found (in the right theory) where this relation between the two couplings holds.
In the CW settings the beta function for $\lambda_\phi$ is positive, 
\begin{equation}
  \label{run1lam}
  \frac{d \lambda_{\phi} }{dt} \,= \,
  \beta_{\lambda_\phi} \,= \,
   \frac{1}{4 \pi^2} 
  \left(9 g^{4}_{\sst CW} - 3 g^{2}_{\sst CW} \lambda_{\phi}+
    \frac{5}{6} \lambda_{\phi}^2 \right) \,>\,0
      \ , \qquad 
  t:= \log (\mu/M_{\sst UV})\,.
\end{equation}
Thus starting in the UV at a positive initial value of $\tilde\lambda_\phi$
and following the RG running to the IR, one reaches the critical value of the RG scale where $\lambda_\phi$ crosses zero.
Before this happens,
at $\lambda_\phi\ll 1$ and positive, the $\lambda_\phi$ trajectory necessarily crosses the $ \alpha_{\sst CW}^{2}$ trajectory. At the
RG point where the constraint 
\eqref{eq:rad4} is satisfied, the vacuum is reached, the condensate $\langle \phi \rangle$ develops and the coupling freezes. 

Since the couplings run only logarithmically, the vev fixed by the
condition \eqref{eq:rad4} depends exponentially on the coupling
constants.  In fact, in weakly coupled perturbation theory the vev is
naturally generated at the scale which is exponentially smaller than
the UV cutoff $M_{\sst UV}$.  By solving the leading-order
RG-running equation for the coupling $g_{\sst CW}$,
\begin{equation}
  \label{run1}
  \frac{d g_{\sst CW} }{dt} \,= \, \frac{ g_{\sst CW}^3}{48 \pi^2} \,,
\end{equation}
we find
\begin{equation}
  \label{run2}
  \langle \phi\rangle \,=\,M_{\sst UV} \times
  \exp\left[-6 \pi \left(\frac{1}{\alpha_{\sst CW}(\langle \phi\rangle)}-
      \frac{1}{\alpha_{\sst CW}(M_{\sst UV})}\right)\right] \,\ll M_{\sst UV}\,.
\end{equation}

\bigskip

The CW gauge group is spontaneously broken by $\langle \phi \rangle$ and the gauge boson acquires the mass,
$m_{Z'}=g_{\sst CW}|\langle \phi \rangle|$.
The mass of the CW scalar is follows from expanding ~\eqref{Veff1} to the second order in $\varphi = \phi -\langle \phi \rangle$,
\begin{equation}
  \label{phimass}
  m^{2}_{\varphi}\,=\, \frac{3 g^{4}_{\sst CW}}{8\pi^2}\, |\langle
  \phi\rangle |^{2}\,.
\end{equation}
It is parametrically (1-loop)  suppressed relative to the corresponding vector boson mass,
$m^{2}_{\varphi}\,=\, {\frac{3}{2}} \alpha_{\sst CW} m^{2}_{Z'} \, \ll \, m^{2}_{Z'}$.

\bigskip

To adapt the expressions above to a more general CW theory, e.g. the B-L theory where the CW field $\phi$ 
has charge $Q_\phi=2$ under the U(1)$_{\sst CW}$ gauge group, and other matter fields are coupled to the latter with their
$B-L$ charges, or other models, we can re-write \eqref{Veff1}, \eqref{eq:rad4} and \eqref{phimass}
as
\begin{equation}
  \label{Veff11}
  V(\phi)\,=\, \frac{1}{4!}
  \left(
  \lambda_{\phi}(\mu)\,+\,
 \beta_{\lambda_{\phi}}
  \left[\log\left(\frac{\phi}{\mu}\right)
    -\frac{25}{12}\right]\right)\,,
\end{equation}
\begin{equation}
  \label{eq:rad41}
  \lambda_{\phi}(\langle\phi\rangle) \,=\, \frac{11}{6}\,  \beta_{\lambda_{\phi}}(\langle\phi\rangle)\,,
  \end{equation}
\begin{equation}
  \label{phimass0}
 \frac{ m^{2}_{\varphi}}{m_{Z'}^2}\,=\, \frac{3 }{8\pi^2}\, \left(Q_\phi \, g_{\sst CW}\right)^2\,
 |\langle \phi\rangle |^{2}\,.
\end{equation}
Note that Eq.~\eqref{Veff11} is manifestly RG-invariant but it also incorporates the finite contribution 
$C=\frac{25}{12}$, as in \eqref{Veff1}.
The last two equations take these finite contributions into account, e.g.~\eqref{eq:rad41}
originates from $ \lambda_{\phi} \,=\, -\frac{\beta_{\lambda_{\phi}}}{4}+ C\beta_{\lambda_{\phi}}=
\frac{11}{6}\,  \beta_{\lambda_{\phi}}$ and both, the log-dependent and the finite effects are important.

\bigskip

Generalising now this simple model to the SM $\times$ U(1)$_{\sst CW}$ theory,
the classical scalar potential 
in conventions of \cite{Englert:2013gz}  reads,
\begin{equation}
  \label{potentialcoupled11}
  V_{\rm cl}(H,\phi)\,=\, \frac{\lambda_{\rm H}}{2}(H^{\dagger} H)^{2}
   \,+\,\frac{\lambda_{\phi}}{4!}|\phi|^{4}
    \,-\, \lambda_{\rm P}(H^{\dagger}H)|\phi|^{2}\,.
\end{equation}
Here $H$ is the usual SM Higgs complex doublet, which in the unitary gauge takes the form $H^T(x)=\frac{1}{\sqrt{2}}(0, v+h(x))$.
The field CW field $\phi$ is normalised as before, and the Higgs is canonically normalised, $|D_\mu H|^2$.
All scalars have vanishing renormalised masses, 
\begin{equation}
  \label{rencond}
  \frac{\partial^{2} V(H,\phi)}{\partial H^{\dagger} \partial H}\bigg|_{H=\phi=0}=0\,,\quad
  \frac{\partial^{2} V(H,\phi)}{\partial \phi^{\dagger} \partial \phi}\bigg|_{H=\phi=0}=0\,.
\end{equation}
Once these conditions 
are enforced at one scale, they hold \cite{Englert:2013gz} at all RG scales in dimensional regularisation 
maintaining classical scale invariance of the model. 

This potential can be re-written as,
\begin{equation}
    \label{potentialcoupled21}
    V_{\rm cl}(H,\phi)\,=\, \frac{\lambda_{\rm H}}{2}\left(|H|^2
      \,-\, \frac{\lambda_{\rm P}}{\lambda_{\rm H}}|\phi|^{2}\right)^2
    \,+\, \frac{\tilde\lambda_\phi}{4!}
    |\phi|^{4}\, , \quad {\rm where} \quad \tilde\lambda_\phi = \lambda_\phi-12 \lambda_{\rm P}^2/\lambda_H \,.
  \end{equation}
 For small positive portal coupling $\lambda_{\rm P}$, the radiative generation of the CW vev proceeds exactly as before
 with the substitution $\lambda_\phi \rightarrow \tilde\lambda_\phi$.

\medskip

 The value of $\langle \phi \rangle$ is given by \eqref{run2}
and it
 induces the Higgs vev required to minimise the potential \eqref{potentialcoupled2}  such that
 $|\langle H \rangle| \,:=\, \frac{v}{\sqrt{2}}\,=\,\sqrt{ \frac{\lambda_{\rm P}}{\lambda_{\rm H}}}\, |\langle \phi \rangle |$ and triggers the 
 electroweak symmetry breaking. The Higgs self-coupling and the portal coupling responsible for the emergence of EWSB 
 are then determined from $\langle \phi\rangle$ through \cite{Englert:2013gz},
 \begin{equation}
  \label{musm3}
 \frac{1}{\lambda_{\rm P}} \, 
  \frac{1}{2}(125~{\rm GeV})^{2}  \, =\, 
  \frac{\lambda_{\rm H}}{\lambda_{\rm P}} \, \frac{(246~{\rm GeV})^{2}}{2} \, =\, 
   |\langle \phi\rangle|^2 \,,
\end{equation}
where we have set $m_{h} = 125~{\rm GeV}$ and $v= 246~{\rm GeV}$.
  
  \bigskip
  
Of course one can argue that the theory under consideration (or for this matter any theory which does not include quantum gravity or in addition has 
Landau poles in the non-asymptotically free couplings, e.g. the hyper charge, even if these Landau poles occur at above the Planck scale) must have
a UV cutoff above which it breaks down. In this case the unknown more microscopic theory above the cutoff, could (or should(?)) contain 
heavy $\sim M_{\rm Pl}$ degrees of freedom which after being integrated out produce large threshold contributions to the Higgs mass. The quadratic cutoff contributions to the Higgs mass in non-dimensional regularisation schemes is an effective realisation
of these physical effects and implies that the masslessness requirement \eqref{potentialcoupled21} is  the fine-tuning. 
The vanishing masses in \eqref{rencond} in this reading correspond to an exact cancellation
of all the $\sim M_{\sst UV}^2$ terms between the bare masses
terms and the counterterms.

One can also take the view that the unknown microscopic theory might after all not generate such large corrections to the Higgs in order
to preserve the classical
scale invariance of its sub-Planckian effective theory. 

The more practical outcome of this discussion is that from the perspective of our classically scale invariant effective theory which should be valid below and 
up to the Planck scale, there are no cancellations between large scales in dimensional regularisation of this theory and no associated 
problems or uncertainties. Technically the SM$\times$CW theory is natural and there is no fine-tuning.

The computational approach we take is to first define the maximal scale above which the theory should not be used, effectively the Planck scale, and in this theory
regulate all integrals using the analytic continuation prescribed by the dimensional regularisation. With that, the maximal UV scale becomes irrelevant, if
one wishes it can be safely taken to infinity, the integrals are convergent in the UV.
No powers of the cutoff scale appear in the integrals, the poles in epsilon are subtracted in the usual way 
in the MS or $\overline{\rm MS}$ scheme and only the logarithms of the RG scale remain.

\bigskip

\bibliographystyle{h-physrev5}

\end{document}